\documentclass[conference]{IEEEtran}
\usepackage{lscape}    
\usepackage{makecell}  

\usepackage{multirow}  
\usepackage{graphicx}  
\usepackage{booktabs}  
\usepackage{amsmath}
\usepackage{tabularx}
\usepackage{amsfonts}
\usepackage{tikz}

\ifCLASSINFOpdf
\else
\fi

\usepackage{soul}
\begin{document}
%
\title{User Identification with LFI-Based Eye Movement Data Using Time and Frequency Domain Features}



%
\author{\IEEEauthorblockN{Süleyman Özdel\IEEEauthorrefmark{1},
Johannes Meyer\IEEEauthorrefmark{2},
Yasmeen Abdrabou\IEEEauthorrefmark{1}, 
Enkelejda Kasneci\IEEEauthorrefmark{1}}
\IEEEauthorblockA{\IEEEauthorrefmark{1}Chair for Human-Centered Technologies for Learning,
Technical University of Munich, Germany}
\IEEEauthorblockA{\IEEEauthorrefmark{2}Corporate Sector Research and Advance Engineering, Robert Bosch GmbH, Renningen, Germany} }



\maketitle

\begin{abstract}

Laser interferometry (LFI)-based eye-tracking systems provide an alternative to traditional camera-based solutions, offering improved privacy by eliminating the risk of direct visual identification. However, the high-frequency signals captured by LFI-based trackers may still contain biometric information that enables user identification. This study investigates user identification from raw high-frequency LFI-based eye movement data by analyzing features extracted from both the time and frequency domains. Using velocity and distance measurements without requiring direct gaze data, we develop a multi-class classification model to accurately distinguish between individuals across various activities. Our results demonstrate that even without direct visual cues, eye movement patterns exhibit sufficient uniqueness for user identification, achieving 93.14\% accuracy and a 2.52\% EER with 5-second windows across both static and dynamic tasks. Additionally, we analyze the impact of sampling rate and window size on model performance, providing insights into the feasibility of LFI-based biometric recognition. Our findings demonstrate the novel potential of LFI-based eye-tracking for user identification, highlighting both its promise for secure authentication and emerging privacy risks. This work paves the way for further research into high-frequency eye movement data.


\end{abstract}



%

\section{Introduction}
Eye-tracking glasses technology is rapidly evolving and becoming more widespread. Currently, videography-based systems (VOG), which typically utilize infrared cameras integrated into glasses, are among the most common eye-tracking setups~\cite{adhanom2023eye,klaib2021eye}. While these systems provide high-resolution gaze tracking, they have several limitations, including sensitivity to lighting conditions, high computational demands, and hardware constraints that limit high-frequency tracking~\cite{fuhl2020tiny,duchowski2017eye}. Alternative methods, such as electrooculography (EOG), also offer high-resolution data, however, with the cost of being intrusive due to the need for skin-contact electrodes~\cite{pleshkov2022comparison}.

To overcome these challenges, laser interferometry (LFI)-based eye-tracking methods have been developed~\cite{meyer2021novel,meyer2023static,meyer2021compact}. Unlike video-based systems, LFI sensors utilize laser interference to measure eye surface velocity and eye-to-sensor distance. LFI-based eye trackers are robust to lighting conditions~\cite{meyer2022highly,meyerrobust2} and 
enable high-frequency eye tracking while using significantly less power. Additionally, they offer greater privacy by eliminating the risk of identifying individuals through their iris patterns. However, although high-frequency velocity and distance signals from LFI sensors are not direct gaze measurements, they still capture unique movement patterns that can be used for user identification. Consistent with traditional eye movement biometrics research literature ~\cite{kasprowski2004eye, lohr2022eye, makowski2021deepeyedentificationlive}, these signals contain eye movement characteristics that could be used for person identification.



From the literature, we can see that eye movement biometrics has been widely studied, with research broadly categorized into feature extraction-based methods. These methods rely on handcrafted features from gaze patterns, and deep learning-based methods, which learn representations directly from raw eye movement data. One of the earliest works in feature-based methods is conducted by Kasprowski et al.~\cite{kasprowski2004eye}, where they applied Cepstrum coefficients from gaze velocity in a jumping dot task, demonstrating the potential of signal processing techniques for biometric identification. Later studies have examined fixation, saccade, and scanpath characteristics, using traditional machine learning techniques to distinguish individuals~\cite{holland2011biometric, george2016score}. Other approaches have explored alternative representations, such as texture-based features extracted using Gabor Wavelet Transform during a visual search task~\cite{li2018biometric}. To better capture the dynamics of eye movement, Harezlak et al.~\cite{harezlak2021biometric} analyzed features like the power spectrum of gaze motion and the largest Lyapunov exponent of the gaze signal. Additionally, the work done by Lohr et al.~\cite{lohr2020metric} applied metric learning with multilayer perceptron (MLP) networks to embed time-domain features from fixations, saccades, and post-saccadic oscillations. While the above studies have shown promising results, the majority focus on time-domain features, with only a few incorporating frequency-domain information~\cite{bednarik2005eye,cuong2012mel}. More significantly, most of the studies rely on datasets collected under strictly controlled laboratory conditions, often involving chin rests, limiting their real-world applicability.  

On the other hand, deep learning methods have been applied to eye movement biometrics, enabling models to learn patterns directly from gaze data. One of the first works in this direction is the work done by Jia et al.~\cite{jia2018biometric}, which introduced GazeNet, a recurrent neural network (RNN) designed for task-independent biometric recognition. Building on this, Abdelwahab et al.~\cite{abdelwahab2022deep} applied deep metric learning with quantile layers and a Wasserstein-based distance metric, improving biometric identification across different modalities. Similarly, Yin et al.~\cite{yin2022effective} investigated spatiotemporal features of eye movement, showing their effectiveness for biometric verification. Moreover, several recent studies have explored different deep-learning approaches. For example, the DeepEyedentification framework~\cite{makowski2021deepeyedentificationlive, jager2020deep} integrates both micro and macro eye movement information using deep convolutional neural networks (CNNs), refining identification accuracy by capturing fine-grained gaze dynamics. Meanwhile, the Eye Know You Too framework~\cite{lohr2022eye, lohr2022eye2} employed a DenseNet-based model to learn distinctive gaze patterns and has also been evaluated on a longitudinal dataset. While recent deep learning methods offer a more flexible alternative by capturing complex patterns, they require large training datasets and significant computational resources, making them impractical for real-time use on resource-constrained wearable eye trackers.

In this work, we explore user identification using LFI-based eye tracking, leveraging raw velocity and distance measurements without relying on gaze estimation or eye movement classification. Unlike previous works, we utilize an LFI-based eye-tracking dataset that includes both velocity and sensor-to-eye distance measurements collected across a diverse range of activities. This dataset comprises not only static tasks such as reading and video watching but also dynamic conditions like walking and cycling, providing a more realistic representation of real-world use cases. To characterize eye movement patterns in LFI-based eye tracking, we extract a comprehensive set of time- and frequency-domain features, incorporating both established features from eye movement biometrics literature and new features to enhance discriminative power. We present results using traditional machine learning classifiers, specifically Linear SVM and LightGBM, which are lightweight and potentially deployable on resource-constrained wearable devices with further optimization. Our approach enables efficient user recognition, demonstrating the feasibility of identifying individuals from raw LFI sensor data in practical scenarios.

\section{Methodology}

\subsection{LFI Sensor Data}

We utilized the dataset from Meyer et al.~\cite{meyer2021cnn}, collected for an activity recognition study using an LFI-based eye-tracking system. Infrared laser sensors embedded in the lower frame of smart glasses tracked eye rotation and depth changes, measuring surface velocity \( v_{\text{eye}} \) and eye-to-sensor distance  \( d_{\text{eye}}\). 
From their dataset, only 10 participants had labeled high-frequency data (1000 Hz), hence, our analysis focuses on these participants, resulting in \~12 hours of data recordings.


The dataset has seven activities, including stationary tasks (talking, reading, problem-solving, watching a video, and typing) and dynamic activities (walking and cycling). As data was collected from two LFI sensors on one eye, it includes surface velocity  (\( v_1, v_2 \)) and distance (\( d_1, d_2 \)) measurements, offering a high-resolution view of eye dynamics. Experimental setup details are in~\cite{meyer2021cnn}.

\subsection{Pre-processing}
We extracted raw velocity and distance signals from the LFI sensors. The velocity components, \( v_1(t) \) and \( v_2(t) \), represent eye surface velocity in different directions, while \( d(t) \) represents the radial displacement between the sensor and the eye. To extract personalized eye kinematic, we compute acceleration and jerk from velocity:
\begin{equation}
a_i(t) = \frac{\mathrm{d} v_i(t)}{\mathrm{d} t}, \quad 
j_i(t) = \frac{\mathrm{d} a_i(t)}{\mathrm{d} t}, \quad 
i \in \{1,2\}.
\end{equation}
To capture motion direction, we compute the tangent of the velocity components, and the absolute velocity magnitude is obtained as:
\begin{equation}
\theta(t) = \arctan\!\bigg(\frac{v_2(t)}{v_1(t)}\bigg), \quad
\|\mathbf{v}(t)\| = \sqrt{v_1(t)^2 + v_2(t)^2}.
\end{equation}
In the distance measurements, to mitigate static biases, we compute the relative distance change as the difference between consecutive distance readings:
\begin{equation}
\Delta d_i(t) = d_i(t) - d_i(t-1), \quad i \in \{1,2\}.
\end{equation}
Overall, ten features are computed from two sensors, including velocity, acceleration, jerk, and relative distance changes, as well as the tangent of velocity components and absolute velocity magnitude derived from both sensors:
\begin{equation}
\mathbf{F}(t) = \bigl[ v_i(t), a_i(t), j_i(t), \theta(t), \|\mathbf{v}(t)\|, \Delta d_i(t) \bigr],
\quad i \in \{1,2\},
\end{equation}
Subsequently, the time-series data is divided into non-overlapping windows of length \( w \), where each window is represented as:
\begin{equation}
\mathcal{W}_k = \{\mathbf{F}(t) \mid t \in [t_k, t_k + w - 1]\}.
\label{eq-preprocessed}
\end{equation}
From each window \( \mathcal{W}_k \), both time and frequency domain features are extracted to characterize eye movement dynamics.



\subsection{Feature Extraction}
To capture the complexity of eye movement dynamics, we extract a comprehensive set of features from both the time and frequency domains. Each feature is computed separately for each component of the feature vector \(\mathbf{F}(t)\), defined in Equation~\ref{eq-preprocessed}, within each non-overlapping time window \(\mathcal{W}_k\). In the equations below, \(x(t)\) generically denotes a single component of \(\mathbf{F}(t)\).
\subsubsection{Time-Domain Features}
Time-domain features provide information about the statistical properties, dynamic variations, and complexity of signals, enabling a comprehensive characterization of eye movement patterns. First, we compute basic statistical measures such as mean, variance, skewness, and kurtosis to describe the central tendency, dispersion, and asymmetry of the signal. Additionally, other simple statistics, including minimum, maximum, root mean square (RMS), and mean absolute deviation (MAD), were computed to summarize the overall signal amplitude characteristics. Beyond statistical descriptors, the dynamic properties of the signal are important to capture rapid fluctuations, typically observed in eye movements. Peak-to-peak amplitude quantifies the overall range of motion by computing the difference between maximum and minimum values within each window. The zero-crossing rate (ZCR), measuring how frequently the signal changes sign, captures oscillatory behaviors. Signal impulsiveness is assessed via the crest factor, defined as the ratio of the peak amplitude to the RMS value.

To analyze short-term temporal dependencies, we compute lag-1 autocorrelation, capturing consistency or randomness between consecutive signal values:
\begin{equation}
R_1 = \frac{\sum_{t=1}^{w-1} (x(t) - \mu)(x(t+1) - \mu)}{\sum_{t=1}^{w} (x(t) - \mu)^2}.
\end{equation}
Longer-range temporal dependencies are assessed using the Hurst exponent \(H\), which quantifies self-similarity and persistence of fluctuations. The Hurst exponent is computed using Detrended Fluctuation Analysis (DFA)~\cite{peng1995quantification,hardstone2012detrended}:
\begin{equation}
F(n) = \sqrt{\frac{1}{N} \sum_{t=1}^{N} [y(t) - y_n(t)]^2} \propto n^{H},
\end{equation}
where \(F(n)\) is the fluctuation magnitude at scale \(n\), \(y(t)\) is the integrated signal, given by \( y(t) = \sum_{i=1}^{t}[x(i)-\bar{x}] \), where \(\bar{x}\) is the mean of the signal. The local trend \(y_n(t)\) is estimated by a least-squares polynomial fit within each window of size \(n\). Additionally, following previous works~\cite{harezlak2021biometric,harezlak2017eye}, we compute the Largest Lyapunov Exponent (LLE)~\cite{wolf1985determining}, which measures the rate at which nearby trajectories diverge, indicating the presence of chaos. Using Rosenstein’s method~\cite{rosenstein1993practical}, LLE, $\lambda$, which quantifies the average exponential divergence rate of nearby trajectories, is computed as:
\begin{equation}
\lambda = \frac{1}{i\,\Delta t}\left\langle \ln \frac{d_j(i)}{d_j(0)} \right\rangle_{j},
\end{equation}
where \(d_j(i)\) denotes the distance between the \(j\)th pair of nearest neighbors after \(i\) discrete time step, and \(\langle \cdot \rangle\) represents averaging over all pairs and multiple reference points in time. 

The Teager–Kaiser Energy Operator (TKEO)~\cite{kaiser1993some,maragos1993energy} is utilized to highlight instantaneous amplitude fluctuations characteristic of dynamic eye movements. It is computed as the difference between the squared current amplitude and the product of adjacent samples:
\begin{equation}
\Psi[x(t)] = x(t)^2 - x(t-1)\,x(t+1), \quad t = 2,\dots,w-1.\end{equation} 
Finally, to capture the randomness and complexity of movement patterns, entropy is computed from the normalized histogram of signal values as:
\begin{equation}
H(x) = -\sum_{n=1}^{N} p_n \log p_n,\quad p_n = \frac{c_n}{\sum_{i=1}^{N} c_i},
\end{equation}
where \(c_n\) is the count of occurrences in histogram bin \(n\), and \(p_n\) represents the corresponding probability.

 \subsubsection{Frequency-Domain Features}
Frequency-domain analysis is utilized to examine spectral characteristics and oscillatory patterns in eye movement signals. The power spectral density (PSD)  is computed to capture stationary periodic components using Welch’s method~\cite{welch2003use}, which improves the robustness and stability of spectral estimates by reducing variance :
\begin{equation}
P_{xx}(f) = \frac{1}{MU}\sum_{m=1}^{M}\left|\sum_{n=0}^{N-1} x_m(n)w(n)e^{-j2\pi f n/N}\right|^2,
\end{equation}
where \( x_m(n) \) is the \( m \)-th segment, \( w(n) \) is the window function, \( N \) is segment length, \( M \) is number of segments, and \( U = \frac{1}{N}\sum_{n=0}^{N-1} w^2(n) \) normalizes the window power. From the computed PSD, we derive spectral entropy, which quantifies how the spectral power is distributed across frequency components, reflecting the complexity and randomness in frequency-domain patterns:
\begin{equation}
H_{\text{spec}} = -\sum_{k=1}^{K} S_k \log(S_k), \quad S_k = \frac{P_k}{\sum_{i=1}^{K} P_i},
\end{equation}
where $P_k$ represents power at frequency bin $k$, and $S_k$ is the normalized spectral power. To complement stationary analysis, wavelet-based methods~\cite{mallat1999wavelet} capture transient, non-stationary characteristics through the discrete wavelet transform (DWT):
\begin{equation}
W_{j, k} = \sum_{n} x(n)\,\psi_{j, k}(n), \quad
\psi_{j, k}(n) = \frac{1}{\sqrt{2^j}} \psi\left(\frac{n - 2^j k}{2^j}\right),
\end{equation}
where \(W_{j,k}\) are the wavelet coefficients and \(\psi_{j,k}(n)\) is the scaled and shifted mother wavelet at scale \(j\) and translation \(k\). The DWT decomposes the signal into approximation (low-frequency) and detail (high-frequency) components across multiple resolutions. Statistical features (mean, variance, skewness, and kurtosis) are computed from each set of approximation and detail coefficients. Finally, wavelet entropy is calculated to quantify the complexity of energy distribution across wavelet coefficients at each scale \(j\):
\begin{equation}
E_{j,k} = \frac{|W_{j,k}|^2}{\sum_{i=1}^{K_j}|W_{j,i}|^2}, \quad
H_{\text{wave},j} = -\sum_{k=1}^{K_j} E_{j,k}\log(E_{j,k}),
\end{equation}
where \(K_j\) represents the number of coefficients at scale \(j\).
The extracted features used as input for classification models are summarized in Table~\ref{tab:features}.

\begin{table}[t]
\centering
\caption{Extracted Time- and Frequency-Domain Features}
\label{tab:features}
\renewcommand{\arraystretch}{1.2}
\resizebox{\columnwidth}{!}{
\begin{tabularx}{\columnwidth}{>{\bfseries}p{2.8cm} X}
\toprule
\multicolumn{2}{l}{\textbf{Time-Domain Features}} \\
\midrule
\multirow{1}{*}{\makecell[l]{Statistical}}  
& Mean, variance, skewness, kurtosis, minimum, maximum, root mean square (RMS), mean absolute deviation (MAD). \\
\multirow{2}{*}{\makecell[l]{Dynamic and Amplitude }}  
& Peak-to-peak amplitude, zero-crossing rate (ZCR), and crest factor. \\
 
\multirow{1}{*}{\makecell[l]{Temporal  Dependence}}  
& Lag-1 autocorrelation and Hurst exponent (H). \\
\multirow{2}{*}{\makecell[l]{Complexity and Chaotic }}  
& Largest Lyapunov Exponent (LLE), Teager–Kaiser energy operator (TKEO), and entropy. \\
\midrule
\multicolumn{2}{l}{\textbf{Frequency-Domain Features}} \\
\midrule
\multirow{2}{*}{\makecell[l]{Spectral Features}}  
& Mean and peak power in frequency bands (0.1–1, 1–4, 4–8, 8–12, 12–30, 30–50), and spectral entropy. \\
\multirow{2}{*}{\makecell[l]{Wavelet Features}}  
& Wavelet decomposition statistics (mean, variance, skewness, kurtosis) and wavelet entropy. \\ 
\bottomrule
\end{tabularx}
}
\end{table}

\subsection{Classification Model}
We evaluated multiple traditional machine learning classifiers using the extracted features, including K-Nearest Neighbors, Support Vector Machines (SVM) with linear and RBF kernels, Random Forest, LightGBM, and XGBoost. To ensure a fair comparison, we performed hyperparameter tuning for each model using a validation dataset. Overall, SVM and LightGBM achieved the best performance while remaining computationally efficient. Therefore, we report results for both in the first two training modes, while focusing on LightGBM in the third mode, as it consistently performed better than linear SVM.

\subsection{Analysis}

We evaluated our models in three modes: (M 1) training and testing on a single activity to assess identification within that activity, (M 2) a leave-one-activity-out approach to evaluate generalization, and (M 3) training and testing on the full dataset. For Modes 1 and 3, data was split sequentially: 50\% for training, 10\% for validation, and 40\% for testing, while Mode 2 used 10\% of training data for validation. Modes 1 and 2 used a 1000 Hz sampling rate with a 5-second window, while Mode 3 explored different rates and window sizes.

\section{Results \& Discussion}

Table~\ref{table:table-modes} presents the accuracy, EER, FAR, and FRR values for LightGBM and SVM with a linear kernel in M1 and M2. In general, user identification accuracy is generally higher when the model is trained and tested on data from the same activity (M1). Even for dynamic free-roaming activities such as cycling and walking, the accuracy reaches 83.46\% and 84.55\%, respectively. However, in the more challenging leave-one-activity-out setting (M2), where the model is trained without any data from the specific activity, performance naturally decreases compared to M1, dropping to around 60\% for dynamic activities. In contrast, the decline is much smaller for static activities, and for the reading activity, accuracy even increases to 96.03\% while the EER decreases to 1.06\%. This suggests that incorporating data from diverse activities enhances user identification in the unseen reading activity compared to a model trained with only limited reading activity data. This scenario better reflects real-world applications, where eye-tracking data may be used across diverse and unseen tasks.


%



\begin{table}[t]
\centering
\caption{Performance Metrics for Mode 1 (M1) and Mode 2 (M2) by Activity (all values in \%).}
\label{table:table-modes}
\resizebox{\columnwidth}{!}{%
\begin{tabular}{c c cc cc cc cc}
\toprule
& & \multicolumn{2}{c}{\textbf{Accuracy}} 
& \multicolumn{2}{c}{\textbf{EER}} 
& \multicolumn{2}{c}{\textbf{FAR}}
& \multicolumn{2}{c}{\textbf{FRR}} \\
\cmidrule(lr){3-4}\cmidrule(lr){5-6}\cmidrule(lr){7-8}\cmidrule(lr){9-10}
\textbf{Activity} & \textbf{Model} 
& \textbf{M1} & \textbf{M2} 
& \textbf{M1} & \textbf{M2} 
& \textbf{M1} & \textbf{M2} 
& \textbf{M1} & \textbf{M2} \\
\midrule
\multirow{2}{*}{\textbf{Talk}} 
& SVM      & \textbf{88.13} & \textbf{61.53} & \textbf{3.74} & \textbf{15.87} & \textbf{1.34} & \textbf{4.28} & \textbf{15.30} & \textbf{39.35} \\
& LightGBM & 79.29 & 58.16 & 5.73 & 17.70 & 2.33 & 4.65 & 24.61 & 41.70 \\
\midrule

\multirow{2}{*}{\textbf{Read}} 
& SVM      & 94.43 & 94.64 & \textbf{1.41} & 1.40 & 0.63 & 0.60 & 6.18 & 5.40 \\
& LightGBM & \textbf{94.96} & \textbf{96.03} & 1.65 & \textbf{1.06} & \textbf{0.57} & \textbf{0.45} & \textbf{5.77} & \textbf{4.43} \\
\midrule

\multirow{2}{*}{\textbf{Video}}
& SVM      & \textbf{95.07} & \textbf{93.62} & \textbf{1.48} & \textbf{1.73} & \textbf{0.55} & \textbf{0.71} & \textbf{4.60} & \textbf{6.29} \\
& LightGBM & 93.35 & 92.72 & 2.36 & 2.01 & 0.74 & 0.80 & 6.73 & 7.36 \\
\midrule

\multirow{2}{*}{\textbf{Walk}} 
& SVM      & \textbf{84.55} & \textbf{67.26} & 6.21 & 11.87 & \textbf{1.70} & \textbf{3.59} & \textbf{13.13} & \textbf{31.25} \\
& LightGBM & 83.67 & 59.83 & \textbf{4.12} & \textbf{9.99} & 1.80 & 4.44 & 15.06 & 35.77 \\
\midrule

\multirow{2}{*}{\textbf{Type}} 
& SVM      & \textbf{95.02} & \textbf{78.72} & 1.89 & 6.08 & \textbf{0.57} & \textbf{2.35} & \textbf{6.75} & \textbf{22.87} \\
& LightGBM & 94.36 & 77.18 & \textbf{1.72} & \textbf{5.81} & 0.65 & 2.53 & 7.55 & 22.89 \\
\midrule

\multirow{2}{*}{\textbf{Solve}} 
& SVM      & \textbf{93.32} & \textbf{92.92} & \textbf{1.88} & \textbf{1.88} & \textbf{0.74} & \textbf{0.80} & \textbf{6.08} & \textbf{6.94} \\
& LightGBM & 89.98 & 90.98 & 4.10 & 2.22 & 1.12 & 1.01 & 9.95 & 8.40 \\
\midrule

\multirow{2}{*}{\textbf{Cycle}}
& SVM      & \textbf{83.46} & \textbf{70.06} & \textbf{4.67} & \textbf{12.17} & \textbf{1.86} & \textbf{3.39} & \textbf{19.01} & \textbf{30.79} \\
& LightGBM & 81.10 & 57.96 & 7.93 & 15.30 & 2.14 & 4.76 & 23.67 & 41.97 \\
\bottomrule
\end{tabular}%
}
\end{table}

In Table~\ref{tab:performance_combined_lightgbm}, we present the results for Mode 3, where the first part of all activities is used in the training set. With a 1000~Hz sampling rate and a 5-second window size, it achieves 93.14\% accuracy and a 2.52\% EER, as highlighted in the table. As the sampling rate decreases, accuracy declines from 93.14\% at 1000 Hz to 89.38\% at 500 Hz and 85.31\% at 250 Hz. Importantly, accuracy drops considerably below 250 Hz, with a significant decline to 72.97\% at 100 Hz and further to 55.14\% at 50 Hz, while the EER rises to 17.67\%. These results indicate that for robust person identification in LFI-based settings, a sampling rate of 250 Hz should be maintained to ensure reliable performance with the extracted features. Furthermore, we observe that after a 5-second window, accuracy and EER improvements become marginal, with performance declining beyond 8 seconds. Those large windows over-summarize data, losing important details. In contrast, very small ones (e.g. 0.5 seconds), fail to capture enough information. Both reduce feature discriminability for identification. The last part of Table \ref{tab:performance_combined_lightgbm} presents results for 2-second windows with majority voting across \( N = 3, 4, \) and \( 5 \) consecutive windows. For continuous user identification, it is more effective to apply majority voting across consecutive shorter windows rather than relying solely on larger windows. With a 2-second window size, we achieve an accuracy of 89.07\%. When applying majority voting across 5 consecutive windows, accuracy improves to 96.74\%, with a 1.01\% EER, outperforming a single 10-second window. A similar trend is observed when comparing single 6- and 8-second windows to their respective 3×2 and 4×2-second consecutive window counterparts.  In practical applications, utilizing a multi-window approach can enhance performance by combining results through majority voting or other methods to preserve different levels of detail.

\begin{table}[t]
\centering
\caption{Performance Metrics for Different Window Sizes and Sampling Rates for Mode 3 (all values in \%)}
\begin{tabular}{clcccc}
    \toprule
    \textbf{Parameter} & \textbf{Value} & \textbf{Accuracy} & \textbf{EER} & \textbf{FAR} & \textbf{FRR} \\
    \midrule
        \multirow{5}{*}{\makecell{Sampling \\ Rate (Hz) \\ (5s Window)}}
    & 500 (Hz)& 89.38 & 3.72 & 1.18 & 10.56 \\
    & 250 (Hz)& 85.31 & 5.29 & 1.64 & 14.29 \\
    & 125 (Hz)& 76.70 & 8.83 & 2.60 & 23.35 \\
    & 100 (Hz)& 72.97 & 9.75 & 3.02 & 27.35 \\
    & 50  (Hz)& 55.14 & 17.67 & 5.00 & 44.81 \\
    
    \midrule
    \multirow{7}{*}{\makecell{Window \\ Size (s) \\ (1000 Hz)}}
    & 0.5 (s) & 75.55 & 9.17 & 2.72 & 25.06 \\
    & 1 (s) & 83.19 & 6.16 & 1.87 & 17.15 \\
    & 2 (s)  & 89.07 & 3.96 & 1.22 & 10.91 \\
    & 3 (s)  & 91.20 & 3.35 & 0.98 & 8.85 \\
    & 4 (s) & 92.85  & 2.55  & 0.80 & 7.08 \\
    & \textbf{5 (s)} & \textbf{93.14} & \textbf{2.52} & \textbf{0.77} & \textbf{6.81} \\
    & 6 (s)& 93.46 & 2.22 & 0.75 & 6.71 \\
    & 8 (s)& 95.27 & 1.58 & 0.53 & 4.72 \\
    & 10 (s) & 95.01 & 1.85 & 0.56 & 4.90 \\
    \midrule
    \multirow{3}{*}{\makecell{Majority Voting \\ Consecutive Windows \\ ($N \times$ (s))}}
    & 3x2 (s)& 94.33  & 1.87 & 0.63 &  5.67 \\
    & 4x2 (s)& 95.77  & 1.57 & 0.47 & 4.12  \\
    & 5x2 (s) &96.74  & 1.01 & 0.36  & 0.32  \\

    \bottomrule
\end{tabular}
\label{tab:performance_combined_lightgbm}
\end{table}

While this feature-based approach with lightweight classifiers shows promising results on raw data from the new LFI-based eye tracker, certain limitations must be considered. This study relies on a small dataset of only 10 participants, limiting the generalizability of the findings. Additionally, although we split the data based on time order, it was collected in a single day, restricting the ability to observe changes in participants’ biometrics over time. A more comprehensive evaluation requires multi-session data collection with, for example, a one-month gap to capture biometric variability and dynamics.

\section{Conclusion}
In this study, we demonstrated the feasibility of user identification with the emerging LFI-based eye-tracking technology, employing handcrafted features and a lightweight multi-class classifier (LightGBM). By extracting a comprehensive set of time-domain and frequency-domain features from velocity and distance signals, we achieved accurate classification across various activities. Additionally, our results highlighted the significant influence of sampling rate and window size on identification performance, highlighting the importance of carefully selecting these parameters. Although further improvements in robustness and generalization are necessary for real-world scenarios, the promising performance of the extracted features and a lightweight classifier highlights their suitability for biometric identification on wearable LFI-based eye trackers, without requiring the high computational demands of deep learning. Nonetheless, our findings also emphasize potential privacy risks, indicating a need for privacy-preserving measures~\cite{abdrabou2025gaze,bozkir2023eye}.


\section*{Acknowledgment}

The project is supported by the Chips Joint Undertaking (Chips JU)
and its members, including top-up funding by Denmark, Germany,
Netherlands, Sweden, under grant agreement No. 101139942. 







\newpage
\bibliographystyle{IEEEtran}

%

\end{document}